\documentclass[prb,twocolumn,showpacs,preprintnumbers,amsmath,amssymb]{revtex4}

\usepackage{graphicx}
\usepackage{dcolumn}
\usepackage{bm}
\usepackage{amsmath}

\usepackage[dvips]{color}
\usepackage{ulem}

\begin{document}
\title
{
Andreev Transport through Side-Coupled Double Quantum Dots
}

\author{
Yoichi Tanaka$^1$\cite{YTriken}, Norio Kawakami$^{1,2}$, and Akira Oguri$^3$
}

\affiliation{
$^{1}$Department of Applied Physics, Osaka University, Suita, Osaka 565-0871, Japan\\
$^{2}$Department of Physics, Kyoto University, Kyoto 606-8502, Japan\\
$^{3}$Department of Material Science, Osaka City University, Osaka 558-8585, Japan
}%

\date{\today}

\begin{abstract}
We study the transport through 
side-coupled double quantum dots,
connected to normal and superconducting (SC) leads   
with a T-shape configuration. 
We find, using the numerical renormalization group, 
that the Coulomb interaction 
suppresses SC interference in the side dot, 
and enhances the conductance substantially in the Kondo regime. 
This behavior stands in total contrast to a wide Kondo valley 
seen in the normal transport. 
The SC proximity penetrating into the interfacial dot
pushes the Kondo clouds,
which screens the local moment in the side dot,
towards the normal lead to make the singlet bond long. 
The conductance shows a peak of unitary limit as the cloud expands.
Furthermore, two separate Fano structures 
appear in the gate-voltage dependence of the Andreev transport, 
where a single reduced plateau appears in the normal transport. 
\end{abstract}


\maketitle

\begin{figure}[b]
\vspace{-1.5mm}
\includegraphics[scale=0.5]{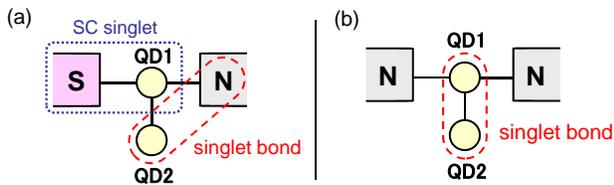}
\caption{
(color online) 
Quantum dots coupled to 
(a) a normal lead (N) and superconductor (S),
and (b) two normal leads. 
QD1 (QD2) is referred to as the interfacial (side) dot.
Dashed line illustrates a dominant singlet pair in the Kondo cloud.
}
\label{modelNDQDScom}
\end{figure}

\section{Introduction}

Observation  of the Kondo effect in a quantum dot (QD) \cite{Gold_Cronen} 
has stimulated researches in the field of quantum transport, and 
recent experimental developments enable one to examine 
the Kondo physics in a variety of systems, 
such as an Aharonov-Bohm (AB) ring with a QD and double quantum dots (DQD).
In these systems multiple paths for electron propagation 
also affect the tunneling currents, and the interference 
causes  Fano-type asymmetric line shapes.

Superconductivity also brings rich and interesting features 
into the quantum transport. Competition between 
superconductivity and the Kondo effect has been reported to be observed  
in  carbon nanotube QD and in semiconductor nanowires.
\cite{Buit:2002,Eich:2007,Sand:2007,Buiz:2007,Grov:2007,SDQDS_exp}
Furthermore, interplay between the Andreev scattering  
and 
the Kondo effect has been studied intensively for
 a QD coupled to a normal (N) lead and superconductor (S),
theoretically 
\cite{Fazio,Schwab,Clerk,Sun,Cuevas,Avishai1,Aono,Krawiec,
Splett,Doma,Avishai2,TanakaNQDS}
and experimentally. \cite{Graber}
So far, however, the Andreev-Kondo physics has been 
discussed mainly for a single dot. 
In this paper, we consider a DQD system with
a T-shape geometry 
as shown in Fig.~\ref{modelNDQDScom},  
and study how multiple paths affect the interplay 
at low temperatures, 
using the numerical renormalization group (NRG) method. \cite{Wilkins}
Golub and Avishai calculated first, to our knowledge, 
the Andreev transport through an AB ring with a QD,\cite{Avishai2} 
in which a similar interference effect is expected. 
However, the underlying Kondo physics in such 
a combination with superconductivity and interference 
is still not fully understood, and is needed to be clarified precisely, 
as measurements are being not impossible.
\cite{SDQDS_exp} 

We find that the Coulomb interaction 
in the side dot (QD2 in Fig.~\ref{modelNDQDScom}) 
suppresses destructive interference typical of
the T-shape geometry, and it enhances substantially 
the tunneling current between the normal and superconducting (SC) leads 
in the Kondo regime. 
This is quite different from the behavior seen in the normal transport 
in the same configuration Fig.~\ref{modelNDQDScom} (b), 
for which the conductance is suppressed, 
and shows a wide minimum called a Kondo valley as 
a result of 
strong interference by the Kondo resonance 
in the side dot.
\cite{Kim,Taka,Corna,YT:2005,Zitko,Karr,Rams,Maruyama2004}
The SC proximity penetrating into the interfacial dot 
(QD1 in Fig.~\ref{modelNDQDScom}) 
causes this stark contrast between the Andreev and normal 
transports. It also changes the Fano line shape 
in the gate-voltage dependence of the conductance. 
Furthermore, we show that the proximity deforms 
the Kondo cloud to make a singlet bond long, 
and it
can be deduced from the Fermi-liquid properties  
of the ground state.

In Sec. \ref{sec:model}, we introduce the model and describe 
the effective Hamiltonian in a large gap limit. 
In Sec. \ref{sec:result}, we show the numerical results 
and clarify the transport properties using the renormalized parameters.  
The Fano structures in the gate-voltage dependence of the conductance
are also discussed. 
A brief summary is given in the last section.

\section{Model} \label{sec:model}

We start with an Anderson impurity connected to 
SC and normal leads, 
\begin{align} 
\,H = H_{DQD} + H_S + H_N + H_{T,S} + H_{T,N},  
\label{Hamipart}
\end{align}
where
\begin{align}
&H_{DQD}
=
\sum_{i=1,2}
\left\{
\left(\varepsilon_{d,i}+\frac{U_i}{2}\right)\left(n_i-1\right)
+\frac{U_i}{2}\left(n_i-1\right)^2
\right\}
\nonumber\\
& \qquad \qquad  
+t\,\sum_{\sigma}\left(d_{1\sigma}^{\dag}d_{2\sigma}^{}+\textrm{H.c.}\right),
\nonumber\\
&H_S
=
\sum_{k,\sigma}\varepsilon _{k}
c_{S,k\sigma}^\dag c_{S,k\sigma}^{}
-\sum_{k}\left(\Delta\, 
c_{S,k\uparrow}^\dag \,c_{S,-k\downarrow}^\dag 
+ \textrm{H.c.}\right),
\nonumber\\
& 
H_N 
= \sum_{k,\sigma}\varepsilon _{k}
c_{N,k\sigma}^\dag c_{N,k\sigma}^{},
\nonumber\\
&H_{T,\nu} = \sum_{k,\sigma} \frac{V_{\nu}}{\sqrt{\mathcal{N}}}
\left(c_{\nu,k\sigma}^\dag d_{1\sigma}^{} + \textrm{H.c.} \right),
 \quad \ \ \nu=S, N.
\label{Hamipart}
\end{align}
$H_{DQD}$ describes the interfacial ($i=1$) and side ($i=2$) dots: 
$\varepsilon_{d,i}$ the energy level, 
$U_{i}$ the Coulomb interaction, 
$n_{i}=\sum_{\sigma}d^{\dag}_{i\sigma}d^{}_{i\sigma}$,
and $t$ the inter-dot hopping matrix element.
$H_{S/N}$ describes the SC/normal lead, and $\Delta$ is 
a $s$-wave BCS gap. 
$V_{S/N}$  is the tunneling matrix element between QD1 and the SC/normal lead. 
We assume that $\Gamma_{S/N}(\varepsilon) \equiv 
\pi V_{S/N}^{2} \sum_k \delta(\varepsilon-\varepsilon_{k})/\mathcal{N}$ 
is a constant independent of the energy $\varepsilon$, 
where $\mathcal{N}$ is the total number of $k$'s in the leads.
Throughout the work,
 we concentrate on a large gap limit $\Delta \to \infty$. 
Then the starting Hamiltonian $H$ can be mapped exactly 
onto a single-channel model, which still 
captures the essential physics of the Andreev reflection 
and makes NRG approach efficient,
\cite{TanakaNQDS,Affleck,Oguri}
\begin{align}
H^\mathrm{eff} =& \ H^\mathrm{eff}_S + H_{DQD} + H_{T,N} + H_N 
\label{Hamieff} \;, \\
H^\mathrm{eff}_S = & \,  
-\Delta_{d1} 
\left(d^{\dag}_{1\uparrow }d^{\dag}_{1\downarrow }+\textrm{H.c.}
\right) \;, 
\label{Hamieff_S} \\
\Delta_{d1} \equiv & \  \Gamma_S \;.
\label{Delta_d1}
\end{align}
Note that at $\Delta \to \infty$  the real and virtual excitations 
towards the continuum states outside the gap 
in the SC lead are prohibited. 
Nevertheless, the proximity from the SC lead to the dot remains finite, 
and it induces a local static pair 
potential $\Delta_{d1}$ ($\equiv \Gamma_S$)  at QD1. 
Furthermore, the current can flow between the SC lead and 
the QD1 via $\Gamma_S$.

\section{Numerical Results} \label{sec:result}

\subsection{$\varepsilon_{d2}$-dependence}

\begin{figure}[t] 
\includegraphics[scale=0.7]{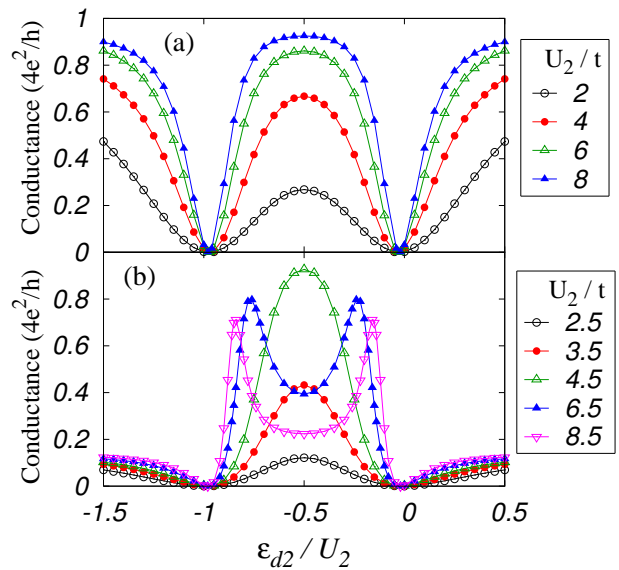}
\caption{
(color online) 
Conductance vs $\varepsilon_{d2}/U_2$ 
for (a) $\Gamma_N=1.0t$ and (b) $\Gamma_N=0.2t$, 
for several side-dot repulsions $U_2$.
The parameters for QD1 are chosen to be $\varepsilon_{d1}=U_1=0$, 
and $\Delta_{d1} = 1.0t$ which is
the local SC gap defined by $\Delta_{d1} \equiv \Gamma_S$. 
}
\label{con-ed2}
\end{figure}

We can calculate the conductance 
at zero temperature as a function of the level position $\varepsilon_{d2}$
of QD2 for different values of $U_2$, 
using the Kubo formula.\cite{Izumida}
In this paper we focus on the Coulomb interaction in the side dot (QD2), 
assuming that $U_{1}=0$ in the following.
The results of the conductance 
are shown in Fig.~\ref{con-ed2} 
for $\varepsilon_{d1}=U_1=0$ and $\Delta_{d1}=1.0t$.  
The coupling to the normal lead is chosen to be 
 (a) $\Gamma_N=1.0t$ and  (b) $\Gamma_N=0.2t$.
The conductance is enhanced for the Kondo regime $-U_2<\varepsilon_{d2} <0$,
where the wide Kondo valley 
appears in the case of the normal transport.
This is a novel phenomenon caused by 
the interplay between superconductivity and the Kondo effect;
the local gap $\Delta_{d1}$ due to the proximity into QD1
leads to the Andreev transport with destructive interference,
but the introduction of $U_2$ suppresses the SC interference 
via QD2, which in turn enhances the conductance.
Note that 
the couplings are symmetric $\Gamma_S =\Gamma_N$ 
for Fig.~\ref{con-ed2} (a), and in this particular case 
the conductance increases with $U_2$ for any $\varepsilon_{d2}$, 
except for the values $\varepsilon_{d2} \simeq -U_2$ and $0.0$.
Outside of the Kondo regime, the side dot is empty or doubly occupied, 
and the interference becomes no longer important.
When the coupling to the normal lead is small $\Gamma_N < 1.0 t$ 
as Fig.~\ref{con-ed2} (b),    
the conductance in the Kondo regime decreases  
after the peak reaches the unitary limit $4e^2/h$. 
This behavior can be related 
to a crossover 
from short-range to long-range Kondo screening 
as illustrated in Fig.~\ref{modelNDQDScom}, 
and is discussed later again.


We examine the behavior 
at the middle point $\varepsilon_{d2}=-U_2/2$ of Fig.~\ref{con-ed2} in detail.
The low-lying energy states show the Fermi-liquid properties,
\cite{TanakaNQDS}
and the conductance can be deduced 
from the renormalized parameters for the quasi-particles
(see Eq.~(\ref{PCstg})).
\begin{figure}[t] 
\includegraphics[scale=0.75]{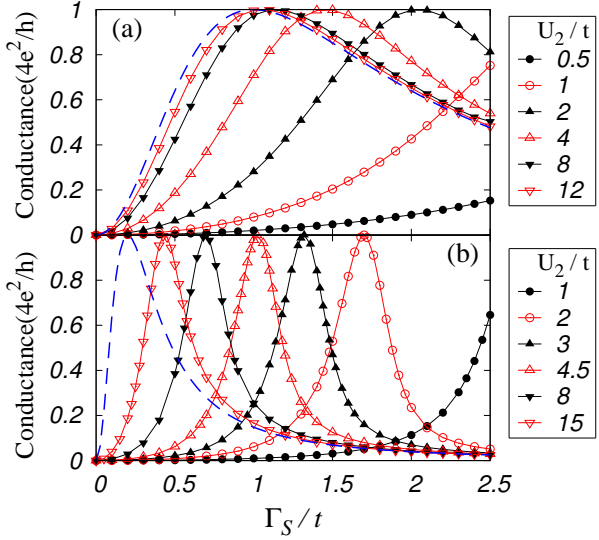}
\caption{
(color online) 
Conductance vs $\Gamma_S/t$ 
at $\varepsilon_{d2}= -U_2/2$ and $\varepsilon_{d1}=U_1=0$  
 for several values of $U_2$. 
 (a) $\Gamma_N=1.0t$ and (b) $\Gamma_N=0.2t$.   
The dashed line is the conductance of a single dot ($t=0$), 
for which the horizontal axis should be interpreted  
as (a) $1.0 \Gamma_S/\Gamma_N$ and (b) $0.2\Gamma_S/\Gamma_N$.
}
\label{con-gs}
\end{figure}
The conductance is plotted as a function of $\Gamma_S/t$ 
in Fig.~\ref{con-gs} 
for (a) $\Gamma_N=1.0t$ and  (b) $\Gamma_N=0.2t$.  
We see that the peak shifts towards  smaller $\Gamma_S$ as $U_2$ increases 
and will coincide 
in the limit of $U_2 \to \infty$ with the dashed line, 
which corresponds to the conductance without the side dot. 
It means that the interference caused by 
the side dot is suppressed completely for large $U_2$,
and in this limit the conductance reaches the unitary limit value 
for the symmetric couplings $\Gamma_S=\Gamma_N$.
The difference in the line shape of Fig.~\ref{con-ed2} (a) and 
that of Fig.~\ref{con-ed2} (b) at fixed $\Gamma_S$ 
reflects the position of the unitary-limit peak in Fig.~\ref{con-gs}.

\subsection{Fermi-liquid properties at $\varepsilon_{d2}=-U_2/2$}
\label{subsec:FL}

In order to clarify the properties of the 
the ground state precisely, 
we consider a special case $\varepsilon_{d2}=-U_2/2$.
Then the Hamiltonian $H^\mathrm{eff}$ 
can be written in terms of the interacting Bogoliubov particles,
which conserve the total charge, 
as shown in  Appendix.\cite{TanakaNQDS}
Consequently, the low-energy states 
can be described by a local Fermi liquid,  
the fixed-point Hamiltonian\cite{Hewson:qp} 
of which can be written in the form 
\begin{align}
& \widetilde{H}_{qp}^{(0)} \  = \  
\widetilde{\Delta}_{d2}\left(d_{2\uparrow }^\dag d_{2\downarrow }^\dag
+\textrm{H.c.}\right )
+
\widetilde{t}\, \sum_{\sigma}\left(d_{1\sigma }^\dag d_{2\sigma}^{}
+\textrm{H.c.}\right)
\nonumber\\
& \qquad \quad 
-\Delta_{d1} \left(d^{\dag}_{1\uparrow }d^{\dag}_{1\downarrow }+\textrm{H.c.}
\right) 
+ H_{T,N}  + H_N .
\label{Hamiqp}
\end{align}
Here, 
$\widetilde{\Delta}_{d2}$ is a local SC gap that emerges in QD2 
via the self-energy correction due to $U_2$, while 
$\Delta_{d1} \equiv \Gamma_S$ as defined in Eq.~(\ref{Delta_d1}) 
is caused by the direct proximity from the SC lead. 
 $\widetilde{t}$ is   
the renormalized value of the inter-dot hopping matrix 
element.
We calculate these parameters 
from the fixed-point of NRG.\cite{Hewson2}
Then, the conductance $G$ and 
a staggered sum $K$ of the pair correlation 
are deduced from the phase shift, $\theta$,  
of the Bogoliubov particles,  
\begin{align}
&
G= \frac{4e^2}{h} \, \textrm{sin}^{2} 2\theta 
\;, \qquad  
K \equiv \sum_{i=1,2} (-1)^{i-1}\, \kappa_{i}
 = \frac{2\theta}{\pi} , 
\label{PCstg} \\
& \tan \theta 
\,\equiv \,
\frac{\widetilde{\Delta}_{d2}\,\Gamma_N}
{\widetilde{t}^2-\widetilde{\Delta}_{d2}\Gamma_{S}} 
, \qquad 
\kappa_{i} \equiv  \langle d_{i\uparrow }^\dag d_{i\downarrow}^\dag
 + d_{i\downarrow}^{} d_{i\uparrow }^{} \rangle . 
\label{theta}
\end{align}

\begin{figure}[t] 
\includegraphics[scale=0.67]{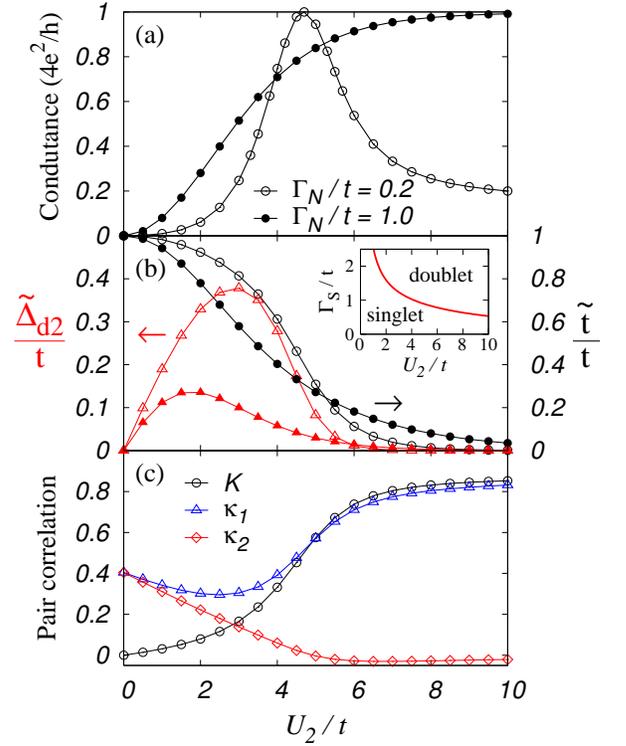}
\vspace{-1mm}
\caption{
(color online) 
Ground state properties at $\varepsilon_{d2}=-U_2/2$: 
 (a) Conductance,  (b)  $\widetilde{t}$, $\widetilde{\Delta}_{d2}$,   
(c) $\kappa_{1}$, $\kappa_{2}$, and $K\equiv \kappa_{1}-\kappa_{2}$. 
We choose $\varepsilon_{d1}=U_1=0$, $\Gamma_S=1.0t$,
and for filled (open) marks $\Gamma_N=1.0t$ ($0.2t$).  
Inset of (b): Phase boundary, between 
 singlet and doublet ground states, 
for an isolated DQD ($\Gamma_N=0$) 
with a finite local SC gap $\Delta_{d1} \equiv \Gamma_S$. 
}
\label{all}
\end{figure}

%
In Fig.~\ref{all}, we show the $U_2$ dependence of 
the ground-state properties at $\varepsilon_{d2}=-U_2/2$
 for $\varepsilon_{d1}=U_1=0$ and $\Gamma_S=1.0t$.
The coupling is chosen to be $\Gamma_N =0.2t$, 
and $1.0t$. 
The conductance for $\Gamma_N < \Gamma_S$ shows 
a peak as a function of $U_2$,
while for $\Gamma_N=\Gamma_S$ it increases simply 
towards the unitary limit.
This corresponds to the difference that we see  
in Fig.~\ref{con-ed2}(a) and (b) at $\varepsilon_{d2}=-U_2/2$.
Figure \ref{all}(b) shows the renormalized parameters 
($\circ,\bullet$) $\widetilde{t}$ 
and ($\vartriangle$,$\blacktriangle$) 
 $\widetilde{\Delta}_2$.
The ratio $\widetilde{t}/t$, 
which equals to 
the square root of the wavefunction renormalization factor $Z$  
(see Appendix),
 decreases monotonically from $1.0$ to $0.0$ with increasing $U_2$,
while the local SC gap $\widetilde{\Delta}_{d2}$ 
becomes large for intermediate values of $U_2$.
The behavior of these Fermi-liquid parameters implies 
that a crossover from weak to strong correlation regimes 
occurs around $U_2 \simeq 4.5t$. 
The nature of 
the crossover can be related
to a level crossing taking place 
in a \textit{molecule} limit $\Gamma_N=0$, 
where QD1 is decoupled from the normal lead.
In this limit, the isolated DQD is described 
by a Hamiltonian $H_{DQD} + H^\mathrm{eff}_S$,
which includes the local SC gap $\Delta_{d1} \equiv \Gamma_S$ at QD1.
The ground state of the \textit{molecule}
is a singlet or doublet depending on  $U_2/t$ and $\Gamma_S/t$,
as shown in the inset of Fig.~\ref{all}(b).
The ground state is a spin-singlet, 
if either $U_2/t$ or $\Gamma_S/t$ is small.
In the opposite case, a spin-doublet becomes the ground state. 
The local moment in this doublet state emerges mainly
at QD2, because 
the correlation at QD1 is small  
in the present situation $\varepsilon_{d1}=U_1=0$.
We see in the phase diagram in Fig.~\ref{all}(b) 
that the transition takes place
in this \textit{molecule} limit
at $U_2 \simeq 4.5t$ for $\Gamma_S=1.0t$,
and it agrees well with the position where 
the conductance peak appears in Fig.~\ref{all}(a).

For finite $\Gamma_N$, conduction electrons can tunnel 
from the normal lead to QD2 via QD1.
However, the SC correlation $\Delta_{d1} \equiv \Gamma_S$ 
tends to make the local state at QD1 a singlet, 
which consists of a linear combination of 
the empty and doubly occupied states.
Thus, for large $\Gamma_S$, the electrons 
at QD1 can not contribute to the screening of the moment at QD2.
In this situation, the Kondo screening 
is achieved mainly by the conduction electrons 
tunneling to QD2 virtually via QD1.
This process is analogous to a superexchange 
mechanism, which can also 
be expected from the Hamiltonian 
written in terms of the Bogoliubov 
particles \eqref{eq:NRG_H_NT_Bogo} in Appendix.
 From these observations we see that 
the conductance peak at $U_2 \simeq 4.5t$ 
in Fig.~\ref{all}(a) reflects the 
crossover from the short-range singlet to long-range 
one due to the superexchange 
screening process (see also Fig.~\ref{modelNDQDScom}) 
 for the Bogoliubov particles.
Note that the peak structure of the conductance 
vanishes for $\Gamma_N =\Gamma_S$. 


The deformation of the Kondo cloud 
can also be deduced from the results shown in Fig.~\ref{all}(c).
This is because the staggered pair correlation $K$ is related  
directly to the scattering phase shift $\theta$ of the Bogoliubov particles, 
by the Friedel sum rule Eq.\ \eqref{eq:Friedel_app} given in Appendix. 
Therefore, the value of $K$ reflects 
the changes occurring in the Kondo clouds.
Particularly, a sudden change observed in $K$ around $U_2 \simeq 4.5t$ 
shows that the phase of the wavefunction shifts 
by $\Delta \theta \simeq 0.4\pi$ during this change.  
This also explains the occurring of the crossover from the short-range 
to long-range screening.  
We can also calculate each correlation function $\kappa_i$ directly 
with NRG based on the definition.
The local SC correlations 
 $\kappa_{1}$ and $\kappa_{2}$ 
have the same value 
in the noninteracting case  $U_2=\varepsilon_{d2}=0$,
and thus in this particular limit 
there is no reduction in  the amplitude of 
the proximity from QD1 to QD2. 
The Coulomb interaction $U_2$ causes the reduction,
as both $\kappa_{1}$ and $\kappa_{2}$ decrease 
for small $U_2$ where the ground state is 
a singlet with a \textit{molecule} character.
For large $U_2$, the SC correlation $\kappa_2$ almost vanishes in QD2,
while $\kappa_{1}$ shows an upturn and approaches the value 
expected for $t=0$.
Therefore, the SC proximity into QD1 is enhanced 
when the Kondo cloud expands to form a long-range singlet.
Then the tunneling current is not interfered much 
by the local moment at QD2, 
and flows almost directly without using the path to the side dot.

%
\subsection{Fano line shape for $\varepsilon_{d1} \neq 0$}
%
\begin{figure}[t] 
\includegraphics[scale=0.7]{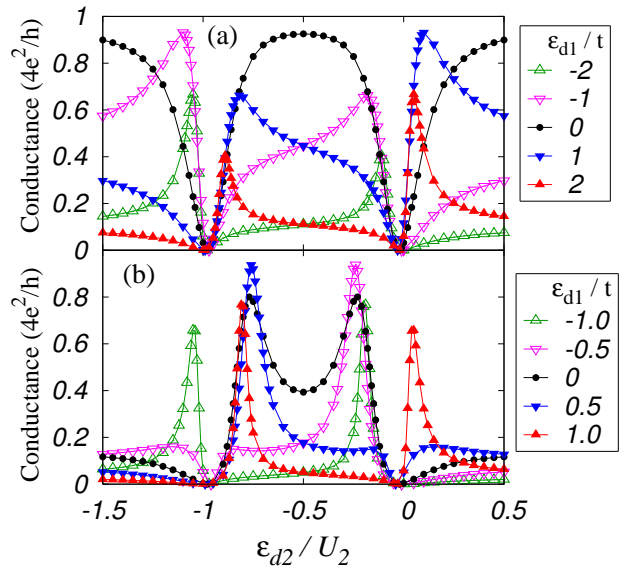}
\vspace{-1mm}
\caption{
(color online) 
Conductance vs $\varepsilon_{d2}$ for several values of $\varepsilon_{d1}$, 
where $U_1=0$ and $\Gamma_S=1.0t$. Other parameters:
(a) $\Gamma_N=1.0t$, $U_2=8.0t$.  
(b) $\Gamma_N=0.2t$, $U_2=6.5t$.  
}
\label{con-ed2_extra}
\end{figure}


So far, we have chosen 
the level of QD1 to be $\varepsilon_{d1}=0$. 
The result obtained for different values of $\varepsilon_{d1}$ 
is plotted as a function of $\varepsilon_{d2}$ 
in Fig.~\ref{con-ed2_extra}. 
We see that two asymmetric Fano structures, each of which consists of 
a pair of peak and dip, emerge at $\varepsilon_{d2}\simeq 0$ 
and $-U_2$ for $\varepsilon_{d1} \neq 0$,
as the Fermi level crosses the energy corresponding to 
the upper and lower levels of the atomic limit.
The conductance peaks become sharper 
for $\Gamma_N < \Gamma_S$ as $|\varepsilon_{d1}|$ increases. 
Maruyama \textit{et al} studied the Fano structure  
in the side-coupled DQD system with two normal leads,\cite{Maruyama2004}
and showed that at low temperatures 
the conductance has only one pair of the peak and dip, 
which are separated widely by a Fano-Kondo plateau 
at $-U_2 \lesssim \varepsilon_{d2}\lesssim 0$. 
This type of the plateau was
known earlier to appear 
in a QD embedded in an AB ring.\cite{Hofstetter}
In contrast, our result shows that the Fano-Kondo plateau vanishes, 
when one of the leads is a superconductor.
This is because the Kondo screening in this case is achieved 
by the long-range singlet bond due to the superexchange process, 
as a result of the competition
between the SC proximity into QD1 
and Coulomb interaction at QD2.

\section{Summary}
We have studied Andreev transport through the
side-coupled DQD with NRG approach.
We have found that the Coulomb interaction 
in the side dot suppresses the destructive
interference effect typical of 
the T-shape geometry, 
and enhances the tunneling current between the normal and SC leads.
This novel phenomenon is caused by the interplay between the
SC correlation and the Kondo effect;
the SC proximity into QD1 pushes the Kondo 
cloud towards the normal lead, 
and the conductance shows a peak of the unitary limit 
as the nature of the singlet changes 
from a short-range to long-range one.
We have also clarified that 
two asymmetric Fano structures appear 
in the gate-voltage dependence of the Andreev transport,
instead of a reduced single Fano-Kondo plateau which 
appears in the Kondo regime of the normal transport.


\begin{acknowledgments}
We thank K.\ Inaba for valuable discussions.
A.O. is also grateful to J.\ Bauer 
and A.\ C.\ Hewson for discussions.
The work is partly supported by a Grant-in-Aid from MEXT Japan
(Grant No.19540338).
Y.T.\ is supported by JSPS Research Fellowships for
Young Scientists.
A.O.\ is supported by JSPS Grant-in-Aid 
for Scientific Research (C).
\end{acknowledgments}

\appendix*

\section{Bogoliubov particles}

The Hamiltonian $H^\mathrm{eff}$ defined in Eq. \eqref{Hamieff} can 
be transformed, at $\varepsilon_{d2}=-U_2/2$,
into the interacting Bogoliubov particles,
which conserve the total charge.
For describing this property briefly, 
we rewrite $H^\mathrm{eff}$ using the logarithmic  
discretization of NRG,\cite{Wilkins}  
%
\begin{align}
&\!\!\!\!
\mathcal{H}_\mathrm{NRG}^\mathrm{eff}  
=
\Lambda^{(N-1)/2}
\left( 
H^\mathrm{eff}_S + H_{DQD} 
+ \mathcal{H}_{T,N} + \mathcal{H}_{N} 
\right)\,,
\nonumber\\
&\!\!\!\!
\mathcal{H}_{T,N} + \mathcal{H}_{N}  
=
\sum_{n=-1}^{N-1} \sum_{\sigma}
t_n \Lambda^{-n/2}
( f_{n+1\sigma}^{\dagger} f_{n\sigma}^{\phantom{\dagger}} +\textrm{H.c.} ) 
.
\label{eq:NRG_SC_cond}
\end{align}
For $n\geq 0$,  the operator $f_{n\sigma}$ describes the conduction 
electron in the normal lead, and  $t_n$ is given by 
\begin{align}
&\!\!\!\!
t_n =  
D\, \frac{1+1/\Lambda}{2}
{ 1-1/\Lambda^{n+1}  
\over  \sqrt{1-1/\Lambda^{2n+1}}  \sqrt{1-1/\Lambda^{2n+3}} 
} \;.
\label{eq:tn}
\end{align}
Here, $D$ is the half-width of the conduction band.
 For the double-dot part, 
we use a notation $f_{-i,\sigma} = d_{i\sigma}$ for $i=1,2$.
Correspondingly, $t_{-1} \equiv \overline{v} \,\Lambda^{-1/2}$ and  
$t_{-2}\equiv t\,\Lambda^{-1}$ with
\begin{align}
&\!\!
  \overline{v}
=\sqrt{ \frac{2\,\Gamma_N D\,A_{\Lambda}}{\pi} },
\quad \ 
A_{\Lambda}=\frac{1}{2} 
\left(\, {1+1/\Lambda \over 1-1/\Lambda }\,\right)
\log \Lambda .
\label{eq:t0}
\end{align}
At $\xi_{d2}\equiv\varepsilon_{d2}+U_2/2 =0$, 
the system has a uniaxial symmetry 
in the Nambu pseudo-spin space,\cite{Yoshihide}
and the Hamiltonian can be simplified by 
the transformation
\begin{align}
&\left[
 \begin{array}{c}
  \gamma_{n\uparrow}^{\phantom{\dagger}} \\
  (-1)^{n-1} \gamma_{n\downarrow}^{\dagger}
 \end{array}
\right]
=
\left[ 
        \begin{array}{cc}
          u  &  -v \\
          v  &  \ u \rule{0cm}{0.5cm}
        \end{array} 
\right]
\left[
 \begin{array}{c}
  f_{n\uparrow}^{\phantom{\dagger}} \\
  (-1)^{n-1} f_{n\downarrow}^{\dagger}
 \end{array}
\right]\;, 
\label{eq:Bogo_B}
\\
&
   u=\sqrt{\frac{1}{2}\left(1+\frac{\xi_{d1}}{E_{d1}}\right)}, \quad
   v=\sqrt{\frac{1}{2}\left(1-\frac{\xi_{d1}}{E_{d1}}\right)} 
\label{eq:Bogo_factor_B} 
\;.
\end{align}
Here, 
$\,E_{d1} \equiv \sqrt{\xi_{d1}^2+|\Delta_{d1}^{\phantom*}|^2 }$,  
$\xi_{d1}\equiv \varepsilon_{d1} + U_{1}/2$,   
and $\Delta_{d1} \equiv \Gamma_S$ as defined 
in Eq.~\eqref{Delta_d1}. 
Then $\mathcal{H}_\mathrm{NRG}^\mathrm{eff}$ 
is transformed into a normal two-impurity Anderson model 
for the Bogoliubov particles 
\begin{align}
&\!\!\!
\mathcal{H}_\mathrm{NRG}^\mathrm{eff}  
=
\Lambda^{(N-1)/2}\Biggl[ 
\,  
E_{d1} 
\left(\widehat{n}_{\gamma, -1} -1
\right)+\sum_{i=1,2} \frac{U_{i}}{2}\left(\widehat{n}_{\gamma,-i}-1\right)^2
\nonumber\\
&
\qquad \quad 
+\sum_{n=-2}^{N-1} \sum_{\sigma}
t_n \Lambda^{-n/2}
\left(
\gamma_{n+1\sigma}^{\dagger}\gamma_{n\sigma}^{\phantom{\dagger}}
+\textrm{H.c.}
\right) \Biggr] .
\label{eq:NRG_H_NT_Bogo}
\end{align}
Here, $\,\widehat{n}_{\gamma i}\equiv \sum_\sigma \gamma_{i\sigma}^{\dagger}
\gamma_{i\sigma}^{\phantom{\dagger}}$, 
and the total number of the Bogoliubov particles,
$\widehat{Q}_{\gamma}\equiv 
\sum_{i=-2}^N \widehat{n}_{\gamma i}$, 
is conserved. 
The equation \eqref{eq:NRG_H_NT_Bogo} 
implies that the low-energy excited states 
can be described by a local Fermi-liquid theory.
This is true also for the original Hamiltonian $H^\mathrm{eff}$, 
and it does not depend on the discretization procedure 
of NRG.\cite{Yoshihide}


To be specific, we assume that $U_{1}=0$ in the following.
In this case, the Bogoliubov particles feel a normal 
impurity potential $E_{d,1}$ at QD1, and this potential 
causes the superexchange mechanism that makes 
the  singlet-bond long range as discussed 
in Sec.~\ref{subsec:FL}.
The retarded Green's function for the Bogoliubov particle
 $\gamma_{-2, \sigma}$ at QD2 takes the form
\begin{align}
G_{\gamma}(\omega) = \frac{1}{\omega - \Sigma_{\gamma}(\omega) 
- \frac{\displaystyle \mathstrut t^2}
{\displaystyle \mathstrut \omega- E^{\phantom{0}}_{d1}
 +i \Gamma_N}}\;, 
\end{align}
where  $\Sigma_{\gamma}(\omega)$
is the self-energy caused by 
the interaction $(U_2/2) \left(\widehat{n}_{\gamma,-2}-1\right)^2$. 
At zero temperature, the asymptotic form of the Green's function 
for small $\omega \simeq 0$ is given by 
\begin{align}
&
G_{\gamma}(\omega) \simeq   \frac{Z}{
\omega  - \widetilde{\Delta}_{d2}
- \frac{\displaystyle \mathstrut \widetilde{t}^{\; 2}}
{\displaystyle \mathstrut \omega- E^{\phantom{0}}_{d1} 
 +i \Gamma_N}
}  \;, 
\label{eq:free_qp}
\\
& \widetilde{\Delta}_{d2}  \equiv   
Z \, \Sigma_{\gamma}(0) , \quad 
\widetilde{t} \equiv  \sqrt{Z} \, t , 
\quad
Z^{-1} \!  \equiv   
1-
\left.\!  
\frac{\partial \Sigma_{\gamma}(\omega)}{\partial \omega}
\right|_{\omega=0} .
\label{eq:Z_til} 
\end{align}
Note that $\widetilde{\Delta}_{d2}$ has 
a finite value even though $\xi_{d2} = 0$,  
because $E_{d1} \neq 0$. 
The value of the parameters  $\widetilde{\Delta}_{d2}$ and $\widetilde{t}$
can be deduced from the fixed point of NRG.\cite{Hewson2}
Then, using the Friedel sum rule for Eq.~\eqref{eq:NRG_H_NT_Bogo}, 
the local charge at the double dot can be calculated
from the phase shift $\theta$ of the Bogoliubov particles, 
%
\begin{align}
&
\left\langle \widehat{n}_{\gamma, -2} \right\rangle + 
\left\langle \widehat{n}_{\gamma, -1} \right\rangle  
\, = \, \,\frac{2}{\pi} \left(\pi-\theta \right) \;, 
\label{eq:Friedel_app}
\\
& 
\qquad 
\theta \equiv 
{\rm tan}^{-1}\left(
\frac{\widetilde{\Delta}_{d2}\Gamma_N}
{\widetilde{t}^2-\widetilde{\Delta}_{d2}E_{d1}}
\right)\, . 
\label{eq:phase_shift_app}
\end{align}
The charge of the Bogoliubov particles corresponds to   
the SC pair correlation for the original electrons $f_{n\sigma}$. 
Specifically for $\xi_{d1}=0$, 
it is transformed into 
the staggered sum $K$ given in Eq.~\eqref{PCstg},
by the inverse transformation of Eq.~\eqref{eq:Bogo_B}.
Similarly, the conductance $G$ can be 
expressed in terms of the phase shift $\theta$.
Furthermore, 
the free quasi-particles corresponding to the Green's function 
given in Eq.~\eqref{eq:free_qp} can be described by a Hamiltonian, 
which is rewritten in terms of the original electron operators 
in Eq.~\eqref{Hamiqp} by the inverse transformation.


\end{document}